\documentstyle[11pt]{article}
\begin{document}
\begin{flushright}
GUTPA/98/10/4\\
\end{flushright}
\vskip .1in
\newcommand{\lapprox}{\raisebox{-0.5ex}{$\
\stackrel{\textstyle<}{\textstyle\sim}\ $}}
\newcommand{\gapprox}{\raisebox{-0.5ex}{$\
\stackrel{\textstyle>}{\textstyle\sim}\ $}}
\newcommand{\lsim}{\raisebox{-0.5ex}{$\
\stackrel{\textstyle<}{\textstyle\sim}\ $}}

\begin{center}

{\Large \bf The Problem of the Quark-Lepton Mass Spectrum }

\vspace{20pt}

{\bf C.D. Froggatt}

\vspace{6pt}

{ \em Department of Physics and Astronomy\\
 Glasgow University, Glasgow G12 8QQ,
Scotland\\}
\end{center}

\section*{ }
\begin{center}
{\large\bf Abstract}
\end{center}

We review some approaches to the quark-lepton mass problem.
The ideas of mass-protection and approximate chiral flavour symmetries
as a framework for resolving the mass hierarchy problem are presented.
Dynamical calculations of the top quark and Higgs particle masses,
based on infra-red fixed points and the so-called multiple point
principle respectively, are discussed. We also consider
mass matrix texture and the neutrino mass puzzle.

\vspace{200pt}

To be published in the Proceedings of the International Workshop on
{\it What comes beyond the Standard Model}, Bled, Slovenia,
29 June - 9 July 1998.

\thispagestyle{empty}
\newpage

\section{Introduction}

The charged fermion masses and mixing angles arise from the Yukawa
couplings, which are arbitrary parameters in the Standard Model (SM).
The masses range over five orders of magnitude, from 1/2 MeV for
the electron to 175 GeV for the top quark. Also the elements of
the quark weak coupling matrix, $V_{CKM}$, range from
$V_{ub} \simeq 0.003$ to $V_{tb} \simeq 1$. This constitutes the
charged fermion mass and mixing problem. It is only the top
quark which has a mass of the order of the electroweak scale
$<\phi_{WS}> = 174$ GeV and a Yukawa coupling constant of order
unity $y_t \simeq 1$. It therefore
seems likely that the top quark mass will be understood
dynamically before those of the other fermions.
All of the other Yukawa couplings are suppressed,
suggesting the existence of physics beyond the SM.   Furthermore the
accumulating evidence for neutrino oscillations provides direct
evidence for physics beyond the SM, in the form of non-zero neutrino
masses.

A fermion mass term essentially represents a transition amplitude between
a left-handed Weyl field $\psi_L$ and a right-handed Weyl field
$\psi_R$. If $\psi_L$ and $\psi_R$ have different quantum numbers,
i.e. belong to inequivalent irreducible representations
of a symmetry group $G$ ($G$ is then called a chiral symmetry),
the mass term is forbidden in the limit of exact $G$ symmetry and
they represent two massless Weyl particles. $G$ thus ``protects'' the
fermion from getting a mass. For example the
$SU(2)_L \times U(1)$ gauge quantum numbers of the left and right-handed
top quark fields are different and the electroweak gauge
symmetry protects the top quark from having a mass,
i.e. the mass term $\overline{t}_Lt_R$ is not gauge invariant.
It is only after the
$SU(2)_L \times U(1)$ gauge symmetry is spontaneously broken that the
top quark gains a mass $m_t=y_t <\phi_{WS}>$, which is consequently
suppressed relative to the presumed fundamental (GUT, Planck...)
mass scale M by the symmetry breaking parameter
$\epsilon = <\phi_{WS}>/M$. The other quarks and leptons have
masses suppressed relative to $<\phi_{WS}>$ and it is natural to
assume that they are protected by further approximately
conserved chiral flavour charges \cite{fn}, as we discuss
further in section 5.

We first consider dynamical calculations of the top quark and
the Higgs particle masses, using Infra-Red Quasi-Fixed Points
in section 2 and the so-called Multiple Point Principle in
section 3. Mass matrix ans\"{a}tze with texture zeros are considered in
section 4.
Finally, the neutrino mass problem is briefly discussed in section 6.
%and section 7 contains our concluding remarks.

\section{Top and Higgs Masses from Infra-red Fixed Point}

The idea that some of the SM mass parameters might be determined as
infra-red fixed point values of renormalisation
group equations (RGEs) was first considered \cite{fn}
some time ago. It was pointed out that the three generation
fermion mass hierarchy does not naturally arise out of the
general structure of the RGEs, although it does seem possible
in special circumstances \cite{abelking}. However it was soon
\cite{pendleton} realised
that the top quark mass might correspond to a fixed point value
or more likely a quasi-fixed point \cite{hill} at the scale
$\mu = m_t$.

The SM quasi-fixed point prediction of the top quark mass
is based on two assumptions:
(a) the perturbative SM is valid up to
some high (e.~g.~GUT or Planck) energy scale
$M \simeq 10^{15} - 10^{19}$ GeV, and
(b) the top Yukawa coupling
constant is large at the high scale $g_{t}(M) \gapprox 1$.
Neglecting the lighter quark masses and mixings, which is
a good approxmation, the SM one loop RGE for the top quark running
Yukawa coupling $g_t(\mu)$ is:
\begin{equation}
16\pi^2\frac{dg_t}{d\ln\mu} = g_t\left(\frac{9}{2}g_t^2 - 8g_3^2
- \frac{9}{4}g_2^2 - \frac{17}{12}g_1^2\right)
\end{equation}
Here the $g_i(\mu)$ are the three SM running gauge coupling constants.
The nonlinearity of the RGEs
then strongly focuses $g_{t}(\mu)$ at the
electroweak scale to its quasi-fixed point value.
The RGE for the Higgs self-coupling $\lambda(\mu)$
\begin{equation}
16\pi^2\frac{d\lambda}{d\ln\mu} =12\lambda^2 +
3\left(4g_t^2 - 3g_2^2 - g_1^2\right)\lambda +
\frac{9}{4}g_2^4 + \frac{3}{2}g_2^2g_1^2 + \frac{3}{4}g_1^4 - 12g_t^4
\label{rgelam}
\end{equation}
similarly focuses $\lambda(\mu)$ towards a
quasi-fixed point value, leading to the SM fixed point predictions \cite{hill}
for the running top quark and Higgs masses:
\begin{equation}
m_{t} \simeq 225\ \mbox{GeV} \quad m_{H} \simeq 250\ \mbox{GeV}
\end{equation}
Unfortunately these predictions are inconsistent with the experimental
running top mass \mbox{$m_{t} \simeq 165 \pm 6$ GeV}.

The corresponding Minimal Supersymmetric Standard Model (MSSM)
quasi-fixed point prediction for the
running top quark mass is
\cite{barger}:
\begin{equation}
m_{t}(m_{t}) \simeq (190\ \mbox{GeV})\sin\beta
\label{mssmfp}
\end{equation}
which is remarkably close to the experimental value for
\mbox{$\tan\beta = 2 \pm 0.5$}.
Some of the soft SUSY breaking parameters are also attracted to
quasi-fixed point values \cite{carena}. For example the trilinear stop
coupling $A_t(m_t) \rightarrow -0.59 m_{gluino}$. For this low
$\tan \beta$ fixed point, there is an
upper limit on the lightest Higgs boson mass:
$m_{h_0} \lsim 100$ GeV. There is also a high $\tan \beta = 60 \pm 5$
fixed point solution \cite{fkm}, corresponding to large Yukawa coupling
constants for the $b$ quark and $\tau$ lepton as well as for
the $t$ quark, sometimes referred to as the Yukawa Unification scenario.
In this case the lightest Higgs boson mass is $m_{h_0} \simeq 120$ GeV.
The origin of the large value of $\tan \beta$ is of course a puzzle
and also SUSY radiative corrections to $m_b$ are then generically large.

\section{Top and Higgs Masses from Multiple Point Principle}

According to the Multiple Point Principle (MPP) \cite{glasgowbrioni},
Nature chooses coupling constant values such that a number of
vacuum states have the same energy density. This principle was
first used in the Anti-Grand Unification Model
(AGUT) \cite{bennett,book},
as a way of calculating the values of the SM gauge coupling constants.
In the Euclidean (imaginary time) formulation, the theory has a phase
transition with the phases corresponding to the degenerate vacua.
The coupling constants then become dynamical, in much the same way
as in baby-universe theory, and take on fine-tuned values
determined by the multiple point. This fine-tuning of the
coupling constants is similar to that of temperature in a
microcanonical ensemble, such as a mixture of ice and water
in a thermally isolated container.

Here we apply the MPP
to the pure SM, which we assume valid up to the Planck scale.
This implies \cite{smtop} that the effective SM
Higgs potential $V_{eff}(|\phi|)$
should have a second minimum
degenerate with the well-known first
minimum at the electroweak scale
$\langle |\phi_{vac\; 1}| \rangle = 174$ GeV.
Thus we predict that our vacuum is barely stable and we
just lie on the vacuum stability curve in the top quark, Higgs
particle mass ($M_t$, $M_H$) plane.
Furthermore we expect the second minimum to be within an
order of magnitude or so of the fundamental scale,
i.e. $\langle |\phi_{vac\; 2}| \rangle \simeq M_{Planck}$.
In this way, we essentially select a particular point on
the SM vacuum stability curve and hence the MPP condition
predicts precise values for the pole masses \cite{smtop}:
\begin{equation}
M_{t} = 173 \pm 5\ \mbox{GeV} \quad M_{H} = 135 \pm 9\ \mbox{GeV}
\end{equation}

\section{Mass Matrix Texture and Ans\"{a}tze}

By imposing symmetries and texture zeros on the fermion
mass matrices, it is possible to obtain testable relations
between the masses and mixing angles.
The best known ansatz for the quark mass matrices
is due to Fritzsch \cite{fritzsch}:
%.
\begin{equation}
M_U =\pmatrix{0  		& C   		& 0\cr
		      C  		& 0   		& B\cr
		      0  		& B   		& A\cr}
\qquad
M_D =\pmatrix{0  		& C^\prime  & 0\cr
		      C^\prime 	& 0   		& B^\prime\cr
		      0  		& B^\prime  & A^\prime\cr}
\end{equation}
It contains 6 complex parameters A, B, C, $A^\prime$, $B^\prime$ and
$C^\prime$, where it is necessary to {\em assume}:
\begin{equation}
|A| \gg |B| \gg |C|, \qquad |A^\prime| \gg |B^\prime| \gg |C^\prime|
\end{equation}
in order to obtain a good fermion mass hierarchy.
Four of the phases can be rotated away by redefining the phases of
the quark fields, leaving just 8 real parameters (the magnitudes of
A, B, C, $A^\prime$, $B^\prime$ and $C^\prime$ and two phases
$\phi_{1}$ and $\phi_{2}$) to reproduce 6 quark masses and
4 angles parameterising $V_{CKM}$. There are thus two relationships predicted
by the Fritzsch ansatz:
\begin{equation}
|V_{us}| \simeq
\left| \sqrt{\frac{m_{d}}{m_{s}}} -
e^{-i\phi_{1}}\sqrt{\frac{m_{u}}{m_{c}}} \right|,
\qquad \qquad |V_{cb}| \simeq
\left| \sqrt{\frac{m_{s}}{m_{b}}} -
e^{-i\phi_{2}}\sqrt{\frac{m_{c}}{m_{t}}} \right|
\label{fritzsch1}
\end{equation}
The first prediction is a generalised version of the relation
$\theta_c\simeq\sqrt{\frac{m_d}{m_s}}$ for the Cabibbo angle,
which originally motivated the two generation Fritzsch ansatz
and is well satisfied experimentally. However the second relationship
cannot be satisfied with a heavy top quark mass
\mbox{$m_{t} > 100$ GeV} and
the original three generation Fritzsch ansatz is excluded by the data.
Consistency with experiment can, for example, be restored by
introducing a non-zero 22 mass matrix element. A systematic analysis of
symmetric quark mass matrices with 5 or 6 texture zeros at the
the SUSY-GUT scale $M_X$ yields five solutions \cite{rrr}.
An example, in which the
non-zero elements are expressed in terms of a small parameter
$\epsilon =\sqrt{\frac{m_c}{m_t}} = 0.058$, is described in
Stech's talk \cite{stech}.

The minimal SU(5) SUSY-GUT relation (using a
Higgs field in the {\bf 5} representation) for the third generation,
$m_b(M_X) = m_{\tau}(M_X)$, is successful. However it cannot be extended
to the first two generations as it predicts $m_d/m_s = m_e/m_{\mu}$,
which fails phenomenologically by an order of magnitude.
This led Georgi and Jarlskog \cite{georgijarlskog}
to introduce an ad-hoc coupling
of the second generation to a Higgs field in the ${\bf \overline{45}}$
representation, giving
mass matrices with the following texture:
\begin{equation}
M_U =\pmatrix{0  		 & C   				& 0\cr
		      C  		 & 0   				& B\cr
		      0  		 & B   				& A\cr}
\quad
M_D =\pmatrix{0  		 & F		& 0\cr
		      F^{\prime} & E   			               & 0\cr
		      0  		 & 0				& D\cr}
\quad
M_E =\pmatrix{0  		& F   				& 0\cr
		      F^{\prime}  		& -3E
	& 0\cr
		      0  		& 0   				& D\cr}
\label{eq:dhransatz}
\end{equation}
and the successful mass relation $m_d/m_s = 9 m_e/m_{\mu}$.
 This ansatz has been developed further in the context
of an SO(10) SUSY-GUT effective operator analysis \cite{anderson}
to give a good fit to all the masses and mixing angles.

\section{Mass Hierarchy from Chiral Flavour Charges}

As we pointed out in section 1, a natural resolution to the charged fermion
mass problem is to postulate the existence of some
approximately conserved chiral charges beyond the SM.
These charges, which we assume to be the gauge quantum numbers
in the fundamental theory beyond the SM, provide selection
rules forbidding the transitions
between the various left-handed and right-handed quark-lepton
states, except for the top quark. In order to generate mass
terms for the other fermion states,
we have to introduce new Higgs fields, which break the
fundamental gauge symmetry group $G$ down to the SM group.
We also need suitable intermediate fermion states to
mediate the forbidden transitions, which we take to be
vector-like Dirac fermions with a mass of order the
fundamental scale $M_F$ of the theory. In this way
effective SM Yukawa coupling constants are generated, which
are suppressed by the appropriate product of Higgs field
vacuum expectation values measured in units of $M_F$.

Consider, for example, an $SMG \times U(1)_f$ model,
obtained by extending the SM gauge group
$SMG = S(U(3) \times U(2)) \simeq SU(3) \times SU(2) \times U(1)$
with a gauged abelian flavour group $U(1)_f$.
$SMG \times U(1)_f$ is broken to SMG
by the VEV of a scalar field $\phi_S$ where
$\langle\phi_S\rangle <  M_F$ and $\phi_S$
carries  $U(1)_f$ charge $Q_f(\phi_S)$ = 1.
Suppose further that $Q_f(\phi_{WS})=0$, $Q_f(b_L)=0$ and $Q_f(b_R)=2$.  Then
it is natural to expect the generation of a $b$ mass of order:
\begin{equation}
\left( \frac{\langle\phi_S\rangle }{M_F} \right)^2\langle\phi_{WS}\rangle
\end{equation}
via the exchange of two $\langle\phi_S\rangle$ tadpoles,
in addition to the usual
$\langle\phi_{WS}\rangle$ tadpole,
through two appropriately charged vector-like
fermion intermediate states \cite{fn}.
We identify
$\epsilon_f=\langle\phi_S\rangle/M_F$
as the $U(1)_f$ flavour symmetry breaking parameter.
In general we expect mass matrix elements of the form:
\begin{equation}
M(i,j) = \gamma_{ij} \epsilon_{f}^{n_{ij}}\langle\phi_{WS}\rangle,
\quad \gamma_{ij} = {\cal O} (1),
\quad n_{ij}= \mid Q_f(\psi_{L_{i}}) - Q_f(\psi_{R_{j}})\mid
\label{eq:mij}
\end{equation}
between the left- and right-handed fermion components.
So the {\em effective\/}
SM Yukawa couplings of the quarks and leptons to the
Weinberg-Salam Higgs field
$y_{ij} = \gamma_{ij}\epsilon_{f}^{n_{ij}}$
can consequently be small even though all
{\em fundamental\/} Yukawa couplings of
the ``true'' underlying theory are of $\cal O$(1).
However it appears \cite{bijnens} not possible to explain the
fermion mass spectrum with an anomaly free set of flavour charges
in an $SMG \times U(1)_f$ model. It is necessary to introduce
SMG-singlet fermions with non-zero $U(1)_f$ charge to
cancel the $U(1)_f^3$ gauge anomaly (as
in $MSSM \times U(1)_f$ models which also use
anomaly cancellation via the Green-Schwarz mechanism \cite{ibanezross})
or by extending the SM gauge group further (as in the
AGUT model \cite{fns}
based on the $SMG^3 \times U(1)_f$ gauge group).

\section{Neutrino Mass and Mixing Problem}

Physics beyond the SM
can generate an effective light neutrino mass term
\begin{equation}
{\cal L}_{\nu-mass} = \sum_{i, j} \psi_{i\alpha}
\psi_{j\beta} \epsilon^{\alpha \beta} (M_{\nu})_{ij}
\end{equation}
in the Lagrangian, where $\psi_{i, j}$ are the Weyl spinors
of flavour $i$ and $j$, and $\alpha, \beta = 1, 2$.
Fermi-Dirac statistics mean that the mass matrix $M_{\nu}$
must be symmetric.
In models with chiral flavour symmetry we typically expect the elements
of the mass matrices to have different orders of magnitude. The charged
lepton matrix is then expected to give only a small contribution
to the lepton mixing. As a result of the symmetry of the neutrino mass
matrix and the hierarchy of the mass matrix elements it is natural
to have an almost degenerate pair of neutrinos, with
nearly maximal mixing\cite{degneut}. This occurs when an off-diagonal
element dominates the mass matrix.

The recent Super-Kamiokande data on the atmospheric neutrino anomaly
strongly suggests large $\nu_{\mu}-\nu_{\tau}$ mixing
with a mass squared difference of order
$\Delta m^2_{\nu_{\mu} \nu_{\tau}} \sim 10^{-3}$ eV$^2$.
Large $\nu_{\mu} - \nu_{\tau}$ mixing is given by the mass matrix
\begin{equation}
M_{\nu} =
\left(
\begin{array}{ccc}\times & \times & \times \\
\times & \times & A \\
\times & A & \times \end{array}\right )
\label{Mnu1}
\end{equation}
where $\times$ denotes small elements and we have
$\Delta m^2_{23} \ll \Delta m^2_{12} \sim \Delta m^2_{13}$,
$\sin^2 \theta_{23} \sim 1$
However, this hierarchy in $\Delta m^2$'s is inconsistent with
the small angle (MSW) solution to the solar neutrino problem,
which requires $\Delta m^2_{12} \sim 10^{-5} \ \mbox{eV}^2$.
Also the theoretically attractive solution \cite{fgn} of the
atmospheric and solar neutrino problems, using maximal
$\nu_e - \nu_{\mu}$ mixing, seems to be ruled out by the zenith
angular distribution of the Super-Kamiokande data.

Hence we need extra structure for the mass matrix such as having
several elements of the same order of magnitude. For example:
\begin{equation}
M_{\nu} =
\left(
\begin{array}{ccc}a & A & B\\
A & \times & \times \\
B & \times & \times \end{array}\right )
\label{Mnu2}
\end{equation}
with $A \sim B \gg a$.
This gives
\begin{equation}
\frac{\Delta m^2_{12}}{\Delta m^2_{23}} \sim \frac{a}{\sqrt{A^2 + B^2}}.
\label{m2rat}
\end{equation}
The mixing is between all three flavours and is given by
the mixing matrix
\begin{equation}
U_{\nu} \sim \left( \begin{array}{ccc}
\frac{1}{\sqrt{2}} & -\frac{1}{\sqrt{2}} & 0\\
\frac{1}{\sqrt{2}} \cos \theta & \frac{1}{\sqrt{2}} \cos \theta &
        -\sin \theta\\
\frac{1}{\sqrt{2}} \sin \theta & \frac{1}{\sqrt{2}} \sin \theta &
        \cos \theta\\
\end{array} \right)
\end{equation}
where $\theta = \tan^{-1} \frac{B}{A}$.
So we have large $\nu_{\mu} - \nu_{\tau}$ mixing with
$\Delta m^2 = \Delta m^2_{23}$, and nearly maximal electron neutrino
mixing with $\Delta m^2 = \Delta m^2_{12}$.
The atmospheric neutrino anomaly requires $\sin^2 2 \theta \gapprox 0.7$
or $1/2 \lsim B/A \lsim 2$. The solar neutrino problem is
explained by vacuum oscillations, although whether it is an
`energy independent' or a `just-so' solution depends on the
value of the mass squared difference ratio in eq.~(\ref{m2rat}).

There is also some difficulty in obtaining the required
mass scale for the neutrinos. In models such as the AGUT the neutrino
masses are generated via super-heavy intermediate fermions in
a see-saw type mechanism. This leads to too small neutrino masses:
\begin{equation} m_{\nu} \lsim \frac{\langle{\phi_{WS}} \rangle^2}{M_F}
\sim 10^{-5}\ \mbox{eV},
\end{equation}
for $M_F = M_{Planck}$ (in general $m_{\nu}$ is also supressed
by the chiral charges). So we need to introduce a new mass scale
into the theory.
Either some intermediate particles with
mass $M_F \lsim 10^{15}\ \mbox{GeV}$, or an $SU(2)$ triplet
Higgs field $\Delta$ with
$\langle \Delta^0 \rangle \sim 1$ eV is required.
Without further motivation the introduction of such particles
is {\em ad hoc}.

%\section{Conclusions}

\section{Acknowledgement}

Financial support of INTAS grant INTAS-93-3316-ext
is gratefully acknowledged.

\end{document}